# Dielectric Dilemma

Robert S. Eisenberg

Department of Applied Mathematics

Illinois Institute of Technology;

Department of Physiology and Biophysics

Rush University Medical Center

Chicago IL

Bob.Eisenberg@gmail.com

February 4, 2019

*File name: Dielectric Dilemma February 3-2 header 2019.docx*

**Abstract**

A dielectric dilemma faces scientists because Maxwell's equations are poor approximations as usually written, with a single dielectric constant. Maxwell's equations are then not accurate enough to be useful in many applications. The dilemma can be partially resolved by a rederivation of conservation of current, where current is defined now to include the 'polarization of the vacuum' $\varepsilon_0\, \partial \mathbf{E}/\partial t$. Conserveration of current becomes Kirchoff's current law with this definition, in the one dimensional circuits of our electronic technology. With this definition, Kirchoff's laws are valid whenever Maxwell's equations are valid, explaining why those laws reliably describe circuits that switch in nanoseconds.

Dielectrics pose a dilemma for scientists. Dielectrics are not well described by the classical equations of electrodynamics, as they are usually written.

The equations of Maxwell [1-5] seem perfectly general in a vacuum, but, in the presence of dielectrics or matter, Maxwell's equations are not general at all. They describe polarization[1] of matter with a severe approximation. They describe polarization by single dielectric constant, a single real number $\geq 1$. This description is nearly universal in textbooks, and in the minds of students, but a single dielectric constant is too approximate to be useful in a technological world where computer circuits switch faster than $10^{-9}$ sec and in a biological world where electrical signals are controlled by atomic movements $< 10^{-10}$m within proteins.

The significant variation of dielectric parameters with time/frequency and conditions has been known a long time [6, 7]. Dielectric parameters have been studied extensively since then [6-36] because polarization is a major determinant of the forces between molecules [37, 38]. The experimental results are remarkably diverse. In solids, dielectric properties depend on time/frequency in every material. In ionic solutions, dielectric properties depend–'without rhyme or reason'–on the type and concentrations of ions [11, 12]. Ionic solutions are of particular importance because all of life occurs in such solutions, as well as a great deal of chemistry. Amidst this diversity, there seems only one uniformity: dielectric properties cannot be described by a single real dielectric constant.

In much higher frequency ranges, of light, for example, dielectric properties determine the refractive index, optical properties, and thus spectra of materials [37], because the polarization of electron orbitals determines how atoms absorb and radiate electromagnetic energy. Spectra are so varied that they are used as fingerprints to identify molecules [37, 39-43]. Spectral properties are as diverse as molecules and obviously cannot be described by a single constant refractive index.

Many interactions of light and materials cannot be described at all by dielectric constants. Dielectric constants are useful only when field strengths are small enough so polarization is a linear phenomenon, independent of field strength. Some of the most interesting applications of electrodynamics involve nonlinear, field dependent

---

[1] We use 'polarization' to mean any charge that depends on the local electric field, for example the charge in an highly polarizable anion like bromide $Br^-$ or the nonuniform charge distribution of $H_2O$ in the liquid state, with its complex time dependent (and perhaps nonlinear) polarization response to the local electric field. The charge or charge densities of alkali metal ions like $Na^+$ and $K^+$ are independent of the local electric field and so are not examples of polarization, but rather can be called permanent or fixed charge.

polarization [44-50].

**The dielectric dilemma** is clear: nonlinearities, spectra, and diverse dielectric behavior cannot be described by a single dielectric constant, but Maxwell's equations use a single dielectric constant, as they are usually written.

When a dielectric is complex, polarization and dielectric behavior need to be described by a functional, and the very form of the Maxwell equations changes. The detailed properties of polarization need to be known and described under the range of conditions in which polarization has significant effects. Polarization is rarely known that well experimentally. Theoretical models or simulations of that scope are also scarce.

Maxwell's equations with a single dielectric constant remain of great importance, however unrealistic the approximation, because that is how they have been taught ([1-5, 51, 52]; [50] is the exception) ever since the equations were formulated [53-56].

Students often remain unaware of the complex properties of the polarization of matter until they become scientists trying to use electrodynamics in applications. As scientists, they face a dielectric dilemma. Too little is often known of polarization to make the Maxwell equations useful in applications demanding precise understanding.

**Conservation of current.** We show here, however, that one property of electrodynamics is independent of polarization, namely conservation of current. That property is in fact enough to ensure that current flow in the one dimensional systems of our electronic technology can be described accurately by the simple, nearly algebraic equations of Kirchoff's current law, without dealing explicitly with field equations and their complexities, including singularities. The accuracy and simplicity of Kirchoff's current law help designers construct robust circuits that switch reliably in $10^{-9}$ sec. Singularities of field equations in three dimensions can increase sensitivity dramatically, making circuits less robust or reliable.

When current $\mathbf{J}_{total}$ is defined to include the 'polarization of the vacuum' $\varepsilon_0\ \partial\mathbf{E}/\partial t$, conservation of current is entirely independent of the polarization and dielectric properties of matter. Lack of experimental knowledge of polarization is unimportant, as far as conservation of current $\mathbf{J}_{total}$ is concerned.

The dielectric dilemma of the Maxwell equations is then partially resolved. Maxwell's equations can be used to analyze current flow in vacuum and matter, without knowledge of polarization. $\mathbf{J}_{total}$ is conserved exactly, in vacuum and dielectrics, indeed in matter in general.

The resolution of the dielectric dilemma is only partial because electrical forces are not determined by currents alone. Forces depend on polarization even if current does not. Where forces are important, Maxwell's equations are useful only if polarization is known in detail. Applications that depend on Kirchoff's current law can take advantage of conservation of current, without knowing anything about polarization or forces at all, as previously mentioned.

**Proof.** Here is the proof of these assertions, following [10, 57-62]. We define current as did Lorrain and Corson [1], p. 276 eq. 6-148.

$$\mathbf{J}_{total} = \mathbf{J} + \overbrace{\varepsilon_0 \frac{\partial \mathbf{E}}{\partial t}}^{\text{Vacuum Displacement Current}} \qquad (1)$$

**E** describes the electric field, with electric constant (permittivity of vacuum) $\varepsilon_0$. **J** describes all the movement of charge that has mass.

**J** includes conduction currents (carried by electrons, holes, or ions, for example). **J** includes all material polarization, no matter how transient and small are the underlying atomic motions. Thus, **J** includes the dielectric current in its classical approximation $(\varepsilon_r - 1)\varepsilon_0 \, \partial \mathbf{E}/\partial t$ as well as all more complex material polarization currents. Here, $\varepsilon_r$ is the relative dielectric coefficient of perfect dielectrics and classical electrodynamics, a single real positive constant $\geq 1$.

**Polarization**. Polarization of two types is important in our derivation. Polarization of matter ('material polarization') arises from the motion of charge, typically from the tiny, transient, reversible motion of electrons in insulators; or the movement of holes and electrons in semiconductors; or the distortion of electron orbitals around atoms or molecules in ionic solutions; or the rearrangement of molecules with asymmetrical distributions of permanent and induced charge, and some organized structure, like water.

**Polarization of the vacuum** $\varepsilon_0 \, \partial \mathbf{E}/\partial t$ is qualitatively different from material polarization and is not a property of matter at all–although it exists in matter–since a vacuum contains no matter. It is not surprising that material models of current in a vacuum $\varepsilon_0 \, \partial \mathbf{E}/\partial t$ are difficult to construct [53-56].

Polarization of the vacuum produces currents $\mathbf{J}_{total}$ that allow electric and magnetic fields to propagate between stars at the velocity of light $c = 1/\sqrt{\varepsilon_0 \, \mu_o}$, determined by the electrical and magnetic constants $\varepsilon_0$, $\mu_o$ and nothing else [1-5, 51, 52].

Polarization of the vacuum is a general property of 'the universe'. It arises from the properties of space-time as described by the theory of relativity [1-5, 51, 52]. The term $\varepsilon_0 \, \partial \mathbf{E}/\partial t$, arises from a special property of charge. Charge is independent of velocity, even at velocities approaching the speed of light, because the Lorentz transformation of charge creates the term $\varepsilon_0 \, \partial \mathbf{E}/\partial t$ (as described in [63]; [2], p. 553; [1], p. 228, eq. 5-110). Mass, length and time all depend on velocity. It does not [2, 63]. Charge is special.

**Maxwell's equations** include Ampere's Law as Maxwell formulated it

$$\frac{1}{\mu_o} \operatorname{curl} \mathbf{B} = \mathbf{J}_{total} = \mathbf{J} + \underbrace{\varepsilon_0 \frac{\partial \mathbf{E}}{\partial t}}_{\substack{\text{Vacuum} \\ \text{Displacement} \\ \text{Current}}} \quad (2)$$

$$\mathbf{J} = \underbrace{(\varepsilon_r - 1)\varepsilon_0 \frac{\partial \mathbf{E}}{\partial t}}_{\substack{\text{Material} \\ \text{Displacement} \\ \text{Current}}} + \mathbf{J}_{everything\ else} \quad (3)$$

**B** describes the magnetic field. The polarization of idealized dielectrics $(\varepsilon_r - 1)\varepsilon_0 \, \partial \mathbf{E}/\partial t$ is isolated from other currents in eq. (2) only for convenience in dealing with the literature. $\mathbf{J}_{everything\ else}$ includes conduction currents (carried by electrons, holes, or ions, for example). $\mathbf{J}_{everything\ else}$ also includes all material polarization, no matter how transient and small are the underlying atomic motions. Thus, $\mathbf{J}_{everything\ else}$ includes the dielectric current in its classical approximation $(\varepsilon_r - 1)\varepsilon_0 \, \partial \mathbf{E}/\partial t$ as well as all the other properties of material polarization and classical conduction.

**Conservation of Current.** The divergence of the curl is zero whenever Maxwell's equations can be used, so conservation of current is as general as Maxwell's equations themselves.

$$\textit{Conservation of Current} \qquad \mathbf{div} \left( \underbrace{\mathbf{J} + \varepsilon_0 \frac{\partial \mathbf{E}}{\partial t}}_{\textit{Current}} \right) = 0 \quad (4)$$

This equation can be solved for **E**. The solution shows that the electric field assumes the value needed to conserve current .

$$\mathbf{E}(x, y, z|t) = -\frac{1}{\varepsilon_0} \int_0^t \mathbf{J}(x, y, z|\tau) \, d\tau \quad (5)$$

**Electric field is an output**. $\mathbf{E}(x, y, z|t)$ is not assumed. It is an output of the analysis. $\mathbf{E}(x, y, z|t)$ is consistent because it is the result of the integration of the Maxwell equation (2) and so depends on 'everything' in a system.

$\mathbf{E}(x, y, z|t)$ produces the polarization of the vacuum $\varepsilon_0\, \partial \mathbf{E}/\partial t$ needed to conserve total current $\mathbf{J}_{total}$, no matter what are the motions of matter that produce $\mathbf{J}$, no matter what are the properties of the material displacement current, including its classical approximation $(\varepsilon_r - 1)\varepsilon_0\, \partial \mathbf{E}/\partial t$.

$\mathbf{E}(x, y, z|t)$ moves charges. The charge movement is a current that combines with $\varepsilon_0\, \partial \mathbf{E}/\partial t$ so Maxwell's equations are satisfied exactly. $\mathbf{E}(x, y, z|t)$ describes the field that produces these charge movements. $\mathbf{E}(x, y, z|t)$ produces the forces and the material currents needed to satisfy Maxwell's equations exactly, when they are combined with the material and vacuum displacement currents.

These words describe the mathematical fact that the Maxwell equations form a consistent theory, in which all variables—including $\mathbf{E}(x, y, z|t)$ and every form of current and flux—satisfy the equations and boundary conditions with one set of unchanging parameters.

**In networks of circuit components**, $\mathbf{E}(x, y, z|t)$ varies dramatically from component to component. The variation is not determined just by the local microphysics of conduction and material polarization but also by the global physics and structure of the network. $\mathbf{E}(x, y, z|t)$ depends on local microphysics, Maxwell equations, and boundary conditions that specify the global properties of the electric field.

The electric field is both global and local. This reality is most vivid in one dimensional networks where components are in series. Currents $\mathbf{J}_{total}$ are equal in every component of a series system, at all times and in all conditions. The current in one component depends on the current in another. The microphysics of conduction in one component does not in itself determine the current flow through that component, despite our local intuition which might suggest otherwise.

Reference [58] discusses this property of series systems in detail, showing how the physics of each component consists of both the local microphysics specific to that component, and also the polarization of the vacuum, the displacement current $\varepsilon_0\, \partial \mathbf{E}/\partial t$. Fig. 3 of [58], and its discussion, show how $\mathbf{E}(x, y, z|t)$ varies in wires, resistors, capacitors, diodes, ionic solutions. $\mathbf{E}(x, y, z|t)$ varies in every component but $\mathbf{J}_{total}$ is always the same, because the components are in series. The currents are equal in the components of the series circuit of Fig. 3 [58], at all times and in all conditions, because the Maxwell equations produce the $\mathbf{E}(x, y, z|t)$ field, the material currents and fluxes, and the 'vacuum' displacement current $\varepsilon_0\, \partial\mathbf{E}/\partial t$ needed to conserve current $\mathbf{J}_{total}$, no matter what are the local microphysics of conduction or polarization [10, 58-60], no matter what the dielectric current is in its classical approximation $(\varepsilon_r - 1)\varepsilon_0\, \partial\mathbf{E}/\partial t$.

**Classical Treatments of Conservation of Current.** The usual derivations [1-5, 51, 52], of conservation of current follow Maxwell [53-56] and use the composite field $\mathbf{D} = \varepsilon_r \varepsilon_0 \mathbf{E}$. The **D** field is a composite because it includes a constitutive property describing the dielectric properties of matter. The **D** field does not have a universal description independent of the properties of matter, because no universal description of the dielectric properties of matter seems possible.

Indeed, the **D** field includes a severe approximation. It treats the dielectric constant $\varepsilon_r$ as a single real constant number, in contradiction to experimental measurements of dielectric properties [6-36], spectra [39-43, 64], and nonlinear polarization [44-50]. When polarization has complex properties, the Maxwell equations as usually written are inadequate. They must be reformulated to describe the polarization that actually exists in the system of interest because that polarization significantly changes the outputs of the equations. Reformulation using the **D** field becomes awkward, if not impossible in many cases, e.g., when polarization is time dependent or nonlinear. In those cases the form of the Maxwell equations changes and, from a mathematical point of view, they become a quite different set of partial differential equations, or integro-differential equations.

**General statements of conservation of current,** like eq. (4), are not easily found in the literature, probably because Maxwell formulated his equations in terms of the **D** field, and others have followed in his footsteps, understandably enough.

The usual derivations of conservation of current describe dielectric properties inaccurately, so conservation of current appears to be inaccurate. Conservation of current appears to be a poor approximation that does not fit experimental data.

The appearance that conservation of current is an approximation is unfortunate because scientists are then reluctant to use the general statement of eq. (4) to understand electrical properties of matter with complex dielectric properties or polarization. Eq. (4) is particularly useful in understanding the physics of current flow in series arrangements like (semiconductor) diodes or ionic channels of biological membranes, where currents are the same everywhere, although the microphysics of conduction of charges is not the same at all.

Conservation of current is not an approximation, as we have just shown. Conservation of current is in fact a general and exact property of the Maxwell equations [10, 60, 62, 65], independent of any properties of matter, true under all conditions. Use of the general principle of conservation of current should allow scientists to better understand and control the flow of current in the circuit boards of our high speed digital devices, or in the crowded confines of protein channels in biological membranes, whether the channels conduct ions, electrons, or charged forms of water like $H^+$ or $H_3O^+$.

**Dielectric dilemma resolved.** The dielectric dilemma posed by traditional formulations of Maxwell's equations is thus partially resolved. Current $\mathbf{J}_{total}$ is conserved. Applications that depend mostly on current conservation, like Kirchoff's current law for electrical networks, can use Maxwell's equations even when dielectric properties are unknown.

The dielectric dilemma is only partially resolved because forces depend on polarization even if current does not. Where forces are important, Maxwell's equations are useful in detail only if polarization is known in detail.

**Engineering Applications.** Engineering applications work reliably in dielectric environments that are rarely known in detail, over a remarkable range of times, starting around $10^{-10}$ sec in the digital circuits of our phones and computers.

Engineering applications use one dimensional systems in which current can be easily measured by (for example) measuring the potential drop across a one ohm (or smaller) resistor placed in series in a branch. All the components in series with that one ohm resistor will have the same current under all conditions and at all times [10, 57-60], independent of the dielectric complexities or conduction mechanism of current in the component. Currents in networks of branches are more complex but networks conserve current as it flows. Conservation of current takes the simple form of the algebraic equations called Kirchoff's current law, if current is defined as $\mathbf{J}_{total}$ [57]. Kirchoff's current law then allows exact analysis of one dimensional systems from the shortest times, because the current law is accurate whenever Maxwell's equations are valid.

It seems hardly a coincidence that so much of our technology uses one dimensional circuits where current can be accurately defined, measured, and controlled, even in nanoseconds. Significant efforts are made to keep the flow of current one dimensional in our high speed technology, particularly in its grounding systems [66], p.787 of [67].

One can hope that current flow in the channels of biology can some day be controlled and exploited (almost) as well as in our semiconductor technology today.